\documentstyle[prl,aps,multicol,epsf]{revtex}

\begin{document}
\draft

\title{Coulomb Blockade Oscillations in the Thermopower of Open Quantum
  Dots.}

\author{A.V.~Andreev$^{1}$ and K.A.~Matveev$^{2}$}
\address{$^{1}$Department of Physics,
  University of Colorado,  CB 390, Boulder, CO 80309\\
  $^{2}$Department of Physics, Duke University, Durham, NC
  27708-0305}

\date{Draft: \today}
\maketitle

\begin{abstract}
  We consider Coulomb blockade oscillations of thermoelectric coefficients
  of a single electron transistor based on a quantum dot strongly coupled to
  one of the leads.  Analytic expression for the thermopower as a function
  of temperature $T$ and the reflection amplitude $r$ in the quantum point
  contact is obtained.  Two regimes can be identified: $T \ll E_C|r|^2$
  and $T \gg E_C |r|^2$, where $E_C$ is the charging energy of the dot.
  The former regime is characterized by weak logarithmic dependence
  of the thermopower on the reflection coefficient, in the latter the
  thermopower is linear in the reflection coefficient $|r|^2$ but depends
  on temperature only logarithmically.
\end{abstract}
\begin{multicols}{2}

Thermoelectric effects in mesoscopic devices have been the subject of
extensive experimental and theoretical research
\cite{Altshuler87,Spivak89,Martinis94,Beenakker92,Staring93,Dzurak97,Matveev99,Molenkamp98}.
The particle-hole asymmetry required for such effects can be strongly
enhanced in these systems as compared to the bulk materials. This and the
small size of such devices make them promising candidates for
technological applications, such as micro-refrigerators~\cite{Martinis94}.
In particular, many experimental and theoretical studies in the last few
years have focused on the thermoelectric properties of quantum dots in the
Coulomb blockade (CB) regime
\cite{Beenakker92,Staring93,Dzurak97,Matveev99,Molenkamp98}.  Most of them
concentrated on the CB oscillations of the thermopower, $S=-\frac{\Delta
  V}{\Delta T}$, where ${\Delta T}$ is the temperature difference across
the dot, and $ \Delta V$ is the voltage necessary to nullify the current.

The theory of the CB oscillations in the thermopower of quantum dots in
the weak tunneling regime was constructed in Ref.~\cite{Beenakker92}.
This theory takes into account only the lowest order tunneling processes,
i.e. the sequential tunneling, and neglects the cotunneling processes.
Its results were in agreement with the experiments of
Ref.~\cite{Staring93}.  Later~\cite{Dzurak97} it became possible to
experimentally access the regime of lower temperatures and stronger
tunneling where the cotunneling processes become dominant. The
theoretical description of this regime was recently given in
Ref.~\cite{Matveev99}.

In very interesting recent experiment \cite{Molenkamp98} the CB
oscillations in the thermopower of a nearly open quantum dot were studied
as a function of the reflection coefficient $|r|^2$ in the contact.  The
setup of these experiments is schematically represented in
Fig.~\ref{fig:1}.  Surprisingly, an initial decrease in the amplitude of
CB oscillations of thermopower with decreasing $|r|^2$ was followed by a
plateau with nearly $|r|$-independent CB oscillations of thermopower. This
saturation was attributed~\cite{Molenkamp98} to the effects of elastic
cotunneling~\cite{Aleiner97}.

The theory of thermopower for the weak tunneling regime developed in
Refs.~\cite{Beenakker92,Matveev99} does not apply to this case.  An
additional motivation for studying the thermoelectric phenomena in such
devices arises from the fact that due to the increased transparency of the
contacts, the open dots are better candidates for micro-refrigerator
devices~\cite{Martinis94} than the closed ones.

In this Letter we present a theory of thermoelectric effects in a quantum
dot in the nearly open regime.  We consider a quantum dot which is coupled
by a tunneling junction to the left lead and by a single channel quantum
point contact (QPC) to the right lead, see Fig.~\ref{fig:1}. The
reflection amplitude in the QPC is assumed to be small, $|r|\ll 1$.  The
mean level spacing $\delta$ in the dot is assumed to be vanishingly
small.  This is a good assumption since experimentally~\cite{Molenkamp98}
$\delta \ll T$.

The previous studies of such systems were devoted to their thermodynamic
and transport properties~\cite{Flensberg93,Matveev95,Furusaki95}.  A
special feature of the thermoelectric power $S$ is that it is sensitive to
the average energy transported by electrons, which in the tunneling
approximation depends on the odd part of the density of states
(DoS) as a function of energy.  Thus the thermoelectric phenomena
represent an {\em   independent} probe of these systems.

In this Letter we find the thermoelectric coefficient $G_T$ of the device
in Fig.~\ref{fig:1} describing the current response $I$ at zero bias,
$\Delta V =0$, to the difference of the temperatures $\Delta T$ between the
two leads: $G_T= \lim \frac{ I}{\Delta T}|_{\Delta V = 0, \Delta T \to
  0}$.  Our main result is the following expression for $G_T$ of the dot:
\begin{eqnarray}
G_T &=& \frac{G_L |r|^2 T}{6 \pi e E_C} \ln{\frac{E_C}{T+\Gamma}} \sin
(2\pi N)
\nonumber \\
&&\times
\int_{-\infty}^{+\infty}\frac{x^2(x^2+\pi^2)}{\left[
    x^2+(\Gamma/T)^2\right]\cosh ^2(x/2)}dx.
\label{eq:result}
\end{eqnarray}
Here $G_L \ll e^2 /h $ is the conductance of the left contact, $e$ is the
absolute value of the electron charge, $E_C$ is the charging energy.  We
have also introduced the energy scale $\Gamma=(8\gamma/\pi^2) E_C |r|^2
\cos ^2 (\pi N)$, which depends on the gate voltage $N$; here $\ln
\gamma={\bf C}\approx 0.5772 \ldots $ is the Euler constant.  The result
(\ref{eq:result}) was obtained with logarithmic accuracy assuming that
$E_C \gg T, \Gamma$.

\narrowtext{
\begin{figure}
\epsfxsize=.8\columnwidth
\hspace{.5em}
\epsfbox{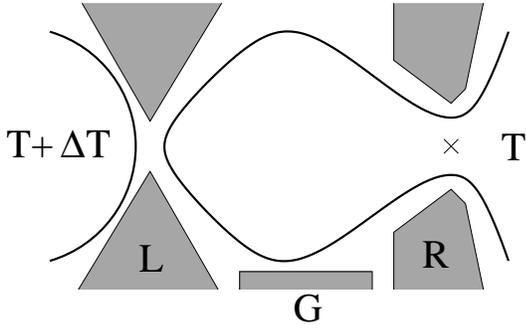}\\[-1ex]
\caption{Schematic drawing of a quantum dot electrostatically
  defined on a surface of a two-dimensional electron gas.  The dot is
  connected to the left lead with temperature $T+\Delta T$ by a tunneling
  junction and to the right lead at temperature $T$ by a single channel
  quantum point contact.  The latter is characterized by a reflection
  amplitude $r$ in the contact and can be thought of as arising from a
  point scatterer depicted by a cross.  The value of $r$ is controlled by
  the voltage on gate R.  The voltage on the central gate G determines the
  optimal electron number $N$ in the dot.}
\label{fig:1}
\end{figure}
}

The thermopower $S=G_T/G$ is then obtained from Eq.~(\ref{eq:result})
using the result of Ref.~\cite{Furusaki95} for the conductance $G$ of the
device shown in Fig.~\ref{fig:1}, which we reproduce here for
completeness:
\begin{equation}
  \label{eq:conductance}
  G=\frac{ G_L \Gamma }{8 \gamma
    E_C}\int_{-\infty}^{+\infty}
  \frac{x^2+\pi^2}{\left[x^2+(\Gamma/T)^2\right]\cosh ^2(x/2)}dx. 
\end{equation}

For the two limiting cases, $T\ll\Gamma$ and $T\gg\Gamma$, we obtain
simplified expressions for the thermopower:
\begin{eqnarray}
  \label{eq:limit}
  S&=& \left\{
  \begin{array}[]{cc}
\frac{64 \gamma |r|^2}{9 \pi^2 e}\ln\frac{E_C}{T} \sin (2\pi N), & {\rm
  for  } \; \;
T\gg \Gamma, \\
\frac{\pi^3 T}{5 e E_C} \ln \frac{E_C}{\Gamma} \tan ( \pi N), & {\rm for  }
\;\;T\ll\Gamma.
  \end{array}\right.
\end{eqnarray}
It is difficult to make a direct comparison of the results
(\ref{eq:result}), (\ref{eq:limit}) with the experiments of
Ref.~\cite{Molenkamp98} since the experimental data were presented in
terms of a fit to the weak-tunneling theory of Ref.~\cite{Beenakker92}.
Nevertheless we want to point out that even without taking into account
the elastic cotunneling effects~\cite{Aleiner97}, in the regime
$T\ll\Gamma$ the thermopower only weakly (logarithmically) depends on the
reflection coefficient, which is consistent with the observation in
Ref.~\cite{Molenkamp98} of the thermopower virtually independent of the
reflection coefficient.  In the opposite regime, $\Gamma \ll T$, the
thermopower is nearly independent of the temperature, but scales linearly
with the reflection coefficient $|r|^2$ vanishing, as expected, at perfect
transmission.  Note that even at very low temperatures $T\ll E_C |r|^2$,
one still has $T\gg\Gamma$ near half-integer values of the gate voltage
$N$ corresponding to the CB peaks of conductance (\ref{eq:conductance}).
The width $\delta N$ of those regions can be easily found from the
condition $\Gamma(N)\sim T$. Upon substitution into Eq.~(\ref{eq:limit})
it gives the estimate of the amplitude of the CB oscillations of the
thermopower $S_0\sim e^{-1} |r| \sqrt{T/E_C} \ln(E_C/T)$.

It is interesting to point out that in the low temperature regime
$T\ll\Gamma$, when the conductance (\ref{eq:conductance}) shows
\cite{Furusaki95} the temperature dependence $G\propto T^2$ characteristic
of inelastic cotunneling\cite{Averin90}, the thermopower can be expressed
in terms of the logarithmic derivative of the conductance with respect to
the gate voltage $2 E_C N$:
\begin{equation}
\label{eq:mott}
S=\frac{\pi^2 T}{10 e E_C}\ln\left(\frac{E_C}{\Gamma}\right)
\frac{\partial \ln G}{ \partial N}. 
\end{equation} 
This form is analogous to the Cutler-Mott formula~\cite{Mott69} for the
thermopower of a system of non-interacting electrons in a metal, but with
a different coefficient in front of the logarithmic derivative.  A similar
Cutler-Mott type relation holds in the case of weak inelastic cotunneling
\cite{Matveev99}; however, the prefactor of Eq.~(\ref{eq:mott}) contains
an additional large logarithmic factor $\ln(E_C/\Gamma)$.  In the opposite
case of high temperature $T\gg \Gamma$ no expression similar to the
Cutler-Mott formula applies.

Below we present the derivation of the result (\ref{eq:result}).
Following Ref.~\cite{Furusaki95}, the electron transport through the right
QPC can be described by a one-dimensional model amenable to bosonization,
whereas the left contact can be treated in the tunneling approximation.
The Hamiltonian of the dot has the form $\hat{H}=\hat{H}_0 + \hat{H}_R +
\hat{H}_L + \hat{H}_C$, where
\begin{mathletters} 
\label{eq:hamiltonian}
\begin{eqnarray}
  \label{eq:hamiltoniana}
  \hat{H}_0 &=&\sum_{k\alpha}\epsilon_k a^\dagger_{k\alpha} a_{k\alpha}
  +\sum_{p\alpha}\epsilon_p a^\dagger_{p\alpha} a_{p\alpha} \nonumber \\ 
&+&\frac{v_F}{2\pi} \sum_{\alpha} \int \left\{[\nabla \phi_\alpha(x)]^2
  +\pi^2\Pi^2_\alpha (x) \right\}dx ,\\
\label{eq:hamiltonianb}
  \hat{H}_L &=& \sum_{kp\alpha} \left( v_t a^\dagger_{k\alpha} a_{p\alpha} F
+     v_t^* a^\dagger_{p\alpha} a_{k\alpha} F^\dagger\right) ,\\
\label{eq:hamiltonianc}
\hat{H}_R &=& \frac{D}{\pi}|r|\sum_{\alpha}\cos[2\phi_\alpha (0) ] ,\\
\label{eq:hamiltoniand}
\hat{H}_C &=& E_C \left[ \hat{n} + \frac{1}{\pi}\sum_\alpha \phi_\alpha
  (0) -N \right]^2 .
\end{eqnarray}
\end{mathletters}
The operators $\hat{H}_R$, $\hat{H}_L$, and $\hat{H}_C$ describe the
backscattering in the right QPC, tunneling through the left contact, and
the charging energy of the dot, respectively.  In the equations above
$\alpha=\uparrow,\downarrow$ is the spin label, $a_{p\alpha}$ and
$a_{k\alpha}$ are electron annihilation operators in the dot and the left
lead respectively, $D$ is the energy cutoff in the bosonization, and
$\phi_\alpha$ is the bosonization displacement operator describing the
electron transport through the right QPC with $\Pi_\alpha$ being its
conjugate momentum, $[\phi_\alpha (x),\Pi_{\alpha'} (x')]=i\delta (x-x')
\delta_{\alpha,\alpha'}$ (we have put $ \hbar =1$).  The modified form of
the tunneling Hamiltonian in Eq.~(\ref{eq:hamiltonianb}) reflects the fact
that the electron tunneling event changes the electron number $\hat{n}$ in
the dot. This is achieved through the introduction of the charge-lowering
operator $F$~\cite{Furusaki95} which satisfies the commutation relation
$[F,\hat{n}]=F$.  The current operator through the left contact can be
obtained from the equation of motion for the charge operator $\hat{I} = -e
\dot{\hat{n}}=ie [\hat{n},\hat{H}]$. Only $\hat{H}_L$ contributes to this
commutator and gives rise to the following expression for the current
operator
\begin{equation}
  \label{eq:currentop}
  \hat{I}= ie\sum_{kp\alpha}\left( 
  v_t^* a^\dagger_{p\alpha} a_{k\alpha} F^\dagger - v_t a^\dagger_{k\alpha}
  a_{p\alpha} F  \right).  
\end{equation}

We treat the problem in the lowest order in the tunneling Hamiltonian
Eq.~(\ref{eq:hamiltonianb}).  We also assume that the conductance of the
tunneling contact is much less than the conductance quantum, $G_L \ll
e^2/h$.  In this approximation all of the temperature drop happens across
the left contact. We take the temperature of the left lead to be $T+\Delta
T$ and that of the dot and the right reservoir to be $T$.  In the linear
approximation in $\Delta T$ the current $I$ can be expressed through the
tunneling DoS $\nu (\epsilon)$ in the dot as,
\begin{equation}
  \label{eq:thermocurrent}
  G_T=\frac{I}{\Delta T}=\frac{G_L}{4 T^2 e \nu_0}\int_{-\infty}^\infty
  \frac{\nu (\epsilon) \epsilon   d  \epsilon}{\cosh^2 \left( \frac{\beta
        \epsilon}{2}\right)}. 
\end{equation}
Here $\nu_0$ is the DoS in the dot in the absence of interaction,
Eq.~(\ref{eq:hamiltoniand}). 

Thus, technically the problem in the tunneling approximation reduces to
the calculation of the energy-dependent tunneling DoS, $\nu (\epsilon)$.
We note that $G_T$ depends only on the odd (as a function of  energy)
component of DoS, 
whereas the conductance $G$ depends only on the even one. Therefore, as
was mentioned earlier, thermopower measurements represent an independent
test of the theory of Coulomb blockade in nearly open dots developed in
Refs.~\cite{Flensberg93,Matveev95,Furusaki95}. Moreover, in the leading
order in ${\rm max} \{T,\Gamma\}/E_C$ the odd component of the tunneling
DoS vanishes~\cite{Furusaki95}. The thermoelectric coefficient $G_T$ is
small in the the ratio of ${\rm max} \{T,\Gamma\}/E_C$ in comparison to
the conductance $G$. Its calculation requires going beyond the previously
adopted approximations~\cite{Furusaki95} and retaining sub-leading order
in $\epsilon/E_C$ in the tunneling DoS, $\nu (\epsilon)$.

The tunneling DoS in the dot can be expressed as 
\begin{equation}
  \label{eq:dos}
  \nu(\epsilon)=-\frac{1}{\pi} \cosh \frac{\beta \epsilon}{2}
  \int_{-\infty}^{\infty} {\cal G}\left( \frac{\beta}{2}+i t\right)
  \exp(i\epsilon t) dt,
\end{equation}
where ${\cal G}\left(\frac{\beta}{2}+i t\right)$ is 
the  Matsu\-ba\-ra Green
function, 
\begin{equation}
  \label{eq:gf}
  {\cal G}(\tau)=-\sum_{pp'} \langle T_\tau 
a_{p\alpha}(\tau)F(\tau)a^\dagger_{p'\alpha}(0)F^\dagger(0) \rangle ,
\end{equation}
analytically continued to complex time $\tau = \frac{\beta}{2}+it$.  The
angular brackets $\langle \ldots \rangle$ in Eq.~(\ref{eq:gf}) denote the
thermal average.

Because the dynamics of the operators $a_{p\alpha}$ and $F$ are decoupled,
the Green function in Eq.~(\ref{eq:gf}) factorizes into ${\cal
  G}(\tau)=G_0(\tau)K(\tau)$, with $G_0(\tau)=\nu_0 \pi T/\sin(\pi T
\tau)$ being the free electron Green function and $K(\tau) = \langle
T_\tau F(\tau)F^\dagger (0)\rangle$, \cite{Furusaki95}.

Since the operator $F^\dagger(0)$ in $K(\tau)$ changes the value of
$\hat{n}$ from zero to one at $t=0$, and $F(\tau)$ changes it back to zero
at $t=\tau$, the correlator $K(\tau)$ can be rewritten as
\begin{equation}
  \label{eq:correlator}
  K(\tau)=\frac{Z(\tau)}{Z(0)},
\end{equation}
where $Z(\tau)$ is a functional integral over $\phi_\alpha$'s in the
presence of the time-dependent charge $n_\tau (t)=\theta (t)\theta(\tau
-t)$.  Introducing the charge and spin mode variables in the right contact
$\phi_{c,s}(x)=[\phi_\uparrow(x)\pm\phi_\downarrow(x)]/\sqrt2$, we can write
$Z(\tau)$ as
\begin{equation}
  \label{eq:funcint}
  Z(\tau)= \int D[\phi_c,\phi_s] \exp[ -{\cal  S}_C (\tau) - 
           {\cal  S}_{0,c}-{\cal  S}_{0,s} -  {\cal S}_R ].  
\end{equation} 
Here ${\cal S}_{0,c}+{\cal S}_{0,s}$ represents the free electron part of
the action in the absence of backscattering in the QPC, $S_C$ denotes its
charging part, and $S_R$ represents the backscattering in the QPC.
These terms are given by
\begin{mathletters}
  \label{eq:action}
\begin{eqnarray}
  \label{eq:actiona}
{\cal S}_{0,i}= \int_0^\beta dt \int dx \frac{v_F}{2\pi} \left( 
[\nabla \phi_i]^2 +\frac{\dot{\phi}_i^2}{v_F^2}  
\right), \quad i=c,s\\
 \label{eq:actionb}
  {\cal S}_{C}(\tau)=\int_0^\beta dt E_C \left[ n_\tau (t) +
  \frac{\sqrt{2}}{\pi}\phi_c 
  (0,t) -N\right]^2, \\
 \label{eq:actionc}
{\cal S}_R = \int_0^\beta dt \frac{2D}{\pi}|r|\cos[\sqrt{2}\phi_c(0,t) ]
\cos[\sqrt{2}\phi_s (0,t)].
\end{eqnarray}
\end{mathletters}

At frequencies below $E_C$ the fluctuations of the charge mode, $\phi_c
(0,t)$ are suppressed by the charging energy term (\ref{eq:actionb}) and
can be integrated out.  Furthermore, we can evaluate the functional
integral over $\phi_c$ by the saddle point approximation ignoring the
backscattering term, Eq.~(\ref{eq:actionc}).
 
The action ${\cal S}^{\rm sp}(\tau)$ and the value of the charge mode 
$\phi^{\rm sp}_c(0,t)$ at the saddle point are found to be 
\begin{mathletters}
\label{eq:sp}
\begin{eqnarray}
  \label{eq:spa}
  {\cal S}^{\rm sp}(\tau)&=&{\cal S}^{\rm sp}_{C}(\tau)+{\cal S}^{\rm
  sp}_{0,c}  
=\ln \frac{2\gamma E_C
  \sin(\pi T \tau)}{\pi^2 T} ,\\
\label{eq:spb}
 \sqrt{2}\phi^{\rm sp}_c(0,t)&=& \pi [N-n_\tau(t)] + {\cal F}(t) + {\cal
 F}(\tau -t) ,\\   
\label{eq:spc}
 {\cal F}(t)&=&\sum_{n=1}^\infty\frac{\sin(2\pi n Tt )}{n+\frac{E_C}{\pi^2
 T}} . 
\end{eqnarray}
\end{mathletters}
In Eq.~(\ref{eq:spa}) we have assumed that $\tau \gg E_C^{-1}$, which is a
good approximation since we only need $\tau=\beta/2+it$ in
Eq.~(\ref{eq:dos}).  

Averaging the backscattering term (\ref{eq:actionc}) over the
fluctuations of $\phi_c$ we obtain
\[
  \tilde{\cal S}_{R,\tau}\! = \!\sqrt{\frac{8 \gamma E_C D}{\pi^3}} |r|
      \!\int_0^\beta \!\! dt  \cos[\sqrt{2}\phi_c^{\rm sp}(0,t)]
\cos[\sqrt{2}\phi_s (0,t)].
\]
Since the charge modes can only be intergated out at frequencies below the
charging energy, one has to assume that the energy cutoff in the above
action is $D\sim E_C$.  

Equation (\ref{eq:funcint}) can now be written as
\begin{mathletters}
\begin{eqnarray}
  \label{eq:truncfia}
  Z(\tau)&=&{\cal N} e^{-{\cal S}^{\rm sp}(\tau) } Z_s(\tau) ,\\
\label{eq:truncfib}
Z_s(\tau) & = & \int D[\phi_s] \exp\left(  -{\cal
  S}_{0,s} -  \tilde{\cal S}_{R,\tau} \right),
\end{eqnarray}
\end{mathletters}
where ${\cal N}$ is the $\tau$-independent factor which arises from the
integration over the fluctuations about the saddle point and drops out of
$K(\tau)$ in Eq.~(\ref{eq:correlator}).  The correlator $K(\tau)$ in
Eq.~(\ref{eq:correlator}) then factorizes into $K(\tau)=K_\Theta
(\tau)K_F(\tau)$, where
\begin{eqnarray}
  \label{eq:ktheta}
  K_\Theta (\tau)&=& e^{- {\cal S}^{\rm sp}(\tau)} =\frac{\pi^2 T}{2\gamma E_C
  \sin(\pi T \tau)} ,
\end{eqnarray}
and $K_F(\tau)$ is the spin part of the correlator which can be expressed
as 
\begin{eqnarray}
  \label{eq:kf}
  K_F(\tau)&=& Z_s (\tau)/Z_s (0).
\end{eqnarray}

The effective action in Eq.~(\ref{eq:truncfib}) can be re-fermionized
following Refs.~\cite{Matveev95,Furusaki95}. The Hamiltonian in this
representation has the form
\begin{mathletters}
\label{eq:fermham}
\begin{eqnarray}
  \label{eq:fermhama}&&
  \hat{H}= i v_F \int \psi^\dagger(x)\nabla \psi(x) dx 
+ \lambda (t)  \eta [\psi(0) - \psi^\dagger(0)], \\&&
  \label{eq:fermhamb}
\lambda(t)=
\frac{2}{\pi}\sqrt{\gamma v_F E_C}|r| \cos[\sqrt{2}\phi_c^{\rm sp}(t)],
\end{eqnarray}
\end{mathletters}
where $\eta=(c+c^\dagger)$ is a Majorana fermion.

In the limit $T/E_C \to 0$ the functions ${\cal F}(t)$ in
Eqs.~(\ref{eq:spb},\ref{eq:spc}) tend to zero, and to the leading order in
$T/E_C$ can be neglected~\cite{Furusaki95}. Then the time-dependent
coefficient $\lambda(t)$ in Eq.~(\ref{eq:fermham}) becomes $\lambda_0(t)=
\frac{2}{\pi}\sqrt{\gamma v_F E_C}|r| (-1)^{n_\tau(t)}\cos[\pi N]$.  In
this approximation~\cite{Furusaki95} the odd  component of the
tunneling DoS in the dot vanishes, thus nullifying the thermopower.
Therefore we expand $K_F(\tau)$ in Eq.~(\ref{eq:kf}) to first order in
$\delta \lambda (t) = \lambda(t)-\lambda_0(t)$. In the fermion
representation (\ref{eq:fermham}) we obtain for the linear in $\delta
\lambda (t)$ correction to $K_F(\tau)$
\begin{mathletters}
  \label{eq:deltak}
\begin{eqnarray}
  \label{eq:deltaka}
  \Delta K_F(\tau)&=& \int_0^\beta (-1)^{n_\tau(t)} \delta \lambda (t)
  \Phi(\tau,t) dt, \\
  \label{eq:deltakb}
 \Phi(\tau,t)&=& \langle T_t \eta(\tau) \eta(0)\eta(t)
 [\psi(0,t)-\psi^\dagger(0,t) ]\rangle,  
\end{eqnarray}
\end{mathletters}
where $\langle \ldots \rangle$ denotes the thermal average with the
Hamiltonian (\ref{eq:fermham}) with $\lambda=\frac{2}{\pi}\sqrt{\gamma v_F
  E_C}|r| \cos[\pi N]$ independent of time $t$.

The average in Eq.~(\ref{eq:deltakb}) can be evaluated with the aid of
Wick theorem. It is not difficult to show that the thermopower is an odd
function of $N$. We therefore need only to retain the odd in $N$ component
$\Delta_{\rm odd} K_F(\tau)$ of Eq.~(\ref{eq:deltaka}).  Evaluating the
integral in Eq.~(\ref{eq:deltaka}) with logarithmic accuracy in $E_C/{\rm
  max}\{T, \Gamma\}$ we find:
\begin{eqnarray}
  \label{eq:deltakodd}
  \Delta_{\rm odd} K_F(\tau)&=&-\frac{8}{E_C}\sqrt{\frac{\gamma \Gamma
    E_C}{\pi}}|r|\sin\left(\pi N\right)\ln\frac{E_C}{T+\Gamma}\nonumber \\
&&\times \int
\frac{\xi d\xi}{\xi^2+\Gamma^2}\frac{e^{\xi |\tau|}}{e^{\beta \xi}+1}. 
\end{eqnarray}
The upper energy scale $E_C$ in the logarithmic factor originates from the
above mentioned energy cutoff $D\sim E_C$ of the spin excitations.  Using
Eq.~(\ref{eq:deltakodd}) we obtain our main result, Eq.~(\ref{eq:result}).

In conclusion, we have presented a theory of the Coulomb blockade
oscillations of the thermoelectric coefficient $G_T$ and the thermopower
$S$ of quantum dots in the anisotropic nearly open regime in the limit
where the single particle mean level spacing is negligible. Two distinct
regimes can be identified: the one with $\Gamma \gg T$, and the one with
$\Gamma \ll T$. In the former the thermopower is linear in temperature but
is nearly independent of the reflection coefficient in the QPC and can be
expressed in the form of Eq.~(\ref{eq:mott}) analogous to the Cutler-Mott
formula~\cite{Mott69}.  In the latter, the thermopower is linear in the
reflection coefficient $|r|^2$ but depends on the temperature only
logarithmically.

We are grateful to B.L.~Altshuler, C.M.~Marcus, J.M.~Martinis,
L.W.~Molenkamp and B.Z.~Spivak for valuable discussions.  It is our
pleasure to acknowledge the warm hospitality of the Aspen Center for
Physics, the ICTP, Trieste, and the Centre for Advanced Studies in Oslo
where part of this work was performed.  The authors are A.P.~Sloan
Research Fellows.  A.A. is a Packard Research Fellow.  This research was
supported by the NSF Grants No.~DMR-9984002 and DMR-9974435.

\end{multicols}


\begin{references}
  
\bibitem{Altshuler87} A.V.~Anisovich, B.L.~Altshuler, A.G.~Aronov, and
  A.Yu.~Zyuzin, JETP Lett {\bf 45}, 295 (1987).

\bibitem{Spivak89} B.Z.~Spivak and A.Yu.~Zyuzin, Europhys.Lett. {\bf 8},
  669 (1989).

\bibitem{Martinis94} M.~Nahum, T.M.~Eiles, and J.M.~Martinis, Appl. Phys.
  Lett. {\bf 65} 3123 (1994).

\bibitem{Beenakker92} C.W.J.~Beenakker and A.A.M.~Staring, Phys.~Rev.
  {\bf B46}, 9667 (1992).

\bibitem{Staring93} A.A.M.~Staring, L.W. Molenkamp, B.W. Alphenhaar, H.
  van Houten, O.J.A. Buyk, M.A.A. Mabesoone, C.W.J. Beenakker, and C.T.
  Foxon, Europhys. Lett. {\bf 22}, 57 (1993).

\bibitem{Dzurak97} A.S.~Dzurak, C.G. Smith, C.H.W. Barnes, M. Pepper, L.
  Martin-Moreno, C.T. Liang, D.A. Ritchie, and G.A.C. Jones, Phys. Rev.
  {\bf B55}, R10197 (1997).

\bibitem{Matveev99}  M. Turek and K.A.~Matveev, {\em in preparation.}

\bibitem{Molenkamp98} S. M\"oller, H. Buhmann, S.F. Godijn, and L.W.
  Molenkamp, Phys. Rev. Lett. {\bf 81}, 5197 (1998).
  
\bibitem{Aleiner97} I.L.~Aleiner and L.I.~Glazman, Phys. Rev. B, {\bf 57},
  9608 (1998).

\bibitem{Flensberg93} K.~Flensberg, Phys. Rev.~{\bf B48}, 11156 (1993);
  Physica (Amsterdam) {\bf 203B}, 432 (1994).

\bibitem{Matveev95} K.A.~Matveev, Phys. Rev. {\bf B51}, 1743 (1995).
  
\bibitem{Furusaki95} A.~Furusaki and K.A.~Matveev, Phys. Rev. Lett.
  {\bf75}, 709 (1995); Phys. Rev. {\bf B52}, 16676 (1995).

\bibitem{Averin90} D.V.~Averin and Yu.V.~Nazarov, Phys. Rev. Lett. {\bf
    65}, 2446 (1990).

\bibitem{Mott69} M.~Cutler and N.F.~Mott, Phys. Rev. {\bf 181}, 1336 (1969).

\end{references}
\end{document}